\documentclass[twocolumn]{aastex631}

\newcommand{\chas}{CH$\alpha$S}

\usepackage{comment}

\shorttitle{Very Extended Ionized Gas Discovered around NGC 1068}
\shortauthors{Melso et al.}

\graphicspath{{./}{figures/}}

\begin{document}

\title{Very Extended Ionized Gas Discovered around NGC 1068 with the Circumgalactic H$\alpha$ Spectrograph}

\correspondingauthor{Nicole Melso}
\email{nmelso@arizona.edu}

\author[0000-0002-4895-6592]{Nicole Melso}
\affiliation{Steward Observatory, University of Arizona, 933 N. Cherry Avenue, Tucson, AZ 85721, USA}

\author{David Schiminovich}
\affiliation{Department of Astronomy, Columbia University, 550 W. 120th Street MC 5246, New York, NY 10027, USA}
\affiliation{Columbia Astrophysics Laboratory, Columbia University, 550 W. 120th St. MC 5247, New York, NY 10027, USA}

\author[0000-0001-7714-6137]{Meghna Sitaram}
\affiliation{Department of Astronomy, Columbia University, 550 W. 120th Street MC 5246, New York, NY 10027, USA}
\affiliation{Columbia Astrophysics Laboratory, Columbia University, 550 W. 120th St. MC 5247, New York, NY 10027, USA}

\author[0000-0001-8255-7424]{Ignacio Cevallos-Aleman}
\affiliation{Department of Physics, Columbia University, 538 W. 120th Street 704 Pupin Hall MC 5255, New York, NY 10027, USA}
\affiliation{Columbia Astrophysics Laboratory, Columbia University, 550 W. 120th St. MC 5247, New York, NY 10027, USA}

\author[0000-0002-1321-3748]{B\'{a}rbara Cruvinel Santiago}
\affiliation{Department of Physics, Columbia University, 538 W. 120th Street 704 Pupin Hall MC 5255, New York, NY 10027, USA}
\affiliation{Columbia Astrophysics Laboratory, Columbia University, 550 W. 120th St. MC 5247, New York, NY 10027, USA}

\author{Brian Smiley}
\affiliation{Columbia Astrophysics Laboratory, Columbia University, 550 W. 120th St. MC 5247, New York, NY 10027, USA}

\author{Hwei Ru Ong}
\affiliation{Columbia Astrophysics Laboratory, Columbia University, 550 W. 120th St. MC 5247, New York, NY 10027, USA}

\begin{abstract}
We have performed wide-field, ultra-low surface brightness H$\alpha$ emission line mapping around NGC 1068 with the newly commissioned Circumgalactic H$\alpha$ Spectrograph (\chas). NGC 1068 is notable for its active galactic nucleus, which globally ionizes gas in the disk and halo. Line-emitting diffuse ionized gas is distributed throughout the galactic disk and large-scale ionized filaments are found well beyond the disk, aligned with the cone angle of the central jet. We report the discovery of a new Ribbon of ionized gas around NGC 1068 beyond even the known outer filamentary structure, located 20 kpc from the galaxy. The H$\alpha$ surface brightness of this Ribbon is on the order of the bright Telluric lines, ranging from $[4-16]$ R with fainter regions on the order of the sky background continuum. Unlike previous extended emission, the Ribbon is not as well aligned with the current axis of the central jet. It is not associated with any galactic structure or known tidal features in the halo of NGC 1068, though it may originate from a larger distribution of unmapped neutral atomic or molecular gas in the halo. The morphology of the Ribbon emission in H$\alpha$ is correlated with extended UV emission around NGC 1068. H$\alpha$ to UV flux ratios in the Ribbon are comparable to extended emission line ratios in the halos of NGC 5128, NGC 253, and M82. The H$\alpha$ excess in the Ribbon gas suggests ionization by slow-shocks or a mixture of in-situ star formation and photoionization and collisional ionization processes. 

\end{abstract}



\section{Introduction} 
\label{sec:intro}
Circumgalactic gas surrounding low-redshift spiral galaxies is both a repository of gas deposited throughout the galaxy's history and a proximate source of material for future accretion onto the galactic disk. Diffuse circumgalactic gas features, ionized by both local and global sources, may comprise a significant fraction of the missing warm/warm-hot gas in the low-redshift baryon budget \citep{Tumlinson2011, Prochaska2011, Shull2012}.  The search for ultra-faint extended gas around galaxies is likely to discover the remnants of past mergers, gas stripping, and material expelled from the galactic disk (e.g., \cite{Yoshida2002, Veilleux2003, Yoshida2016, Boselli2016, Lokhorst2022}).

Emission line mapping is a way to directly access the properties of extended ionized gas in the local universe. H$\alpha$ emission is a direct probe of the warm ionized gas component and traces many processes of interest. It can be used as a proxy for the instantaneous star formation rate and, in compliment, may trace observational signatures of gas accretion at the disk-halo interface \citep{Held2007, Heitsch2009, Bizyaev2017}. Maps of H$\alpha$ emission in a galaxy can be used to localize bursts of star formation, which are likely connected to strong SN-driven outflows \citep{Veilleux2005, Martin2006, Prusinski2021}. Further into the CGM, H$\alpha$ emission emanates from recombining gas ionized by photons spewed from the galaxy or by the integrated UV background radiation field \citep{Donahue1995, Adams2011, Fumagalli2017}.

The detection of extended H$\alpha$ emission may also trace feedback from a central supermassive black hole or AGN, hinting at past and present activity in the galactic nucleus. Extended emission line regions (EELRs) or narrow-line regions (ENLRs) have been detected around Seyfert galaxies, AGN, radio galaxies and QSOs (e.g., \cite{VanBreugel1985, Lintott2009, Keel2012, Schweizer2013, Neff2015, Bait2019, Venturi2023}). Such detections probe photon-mediated processes, the ionizing luminosity and spectrum of the central engine, or alternatively,  mechanical energy input and shock heating of the extended ISM and CGM.  Recently, statistical samples of EELRs and ENLRs have been uncovered in large surveys and used as `light echoes' to infer the luminosity and activity history of AGN on kilo-yr timescales \citep{Keel2012b, Comerford2017, Keel2024}. At the same time, AGN light up the surrounding gas and can reveal its distribution and properties far into the halo (\cite{Moiseev2023} and references therein).

This paper presents the discovery of a new Ribbon of extended ionized gas around NGC 1068. This feature was discovered with the newly commissioned Circumgalactic H$\alpha$ spectrograph (\chas) described in full in \cite{Melso2022}. 
\chas{} is a fast, wide-field, narrowband integral field spectrograph with a moderate spectral resolution and exceptional grasp. It is capable of mapping ultra-low surface brightness diffuse emission with a superb survey speed that competes with integral field spectrographs on 8–10 m
class telescopes. CH$\alpha$S is deployed on the Hilter 2.4-m telescope at MDM Observatory where it is a facility instrument currently available to the consortium and operating in shared-risk mode. 

This is the debut science paper for CH$\alpha$S, presenting H$\alpha$ emission line mapping around NGC 1068 (See Figure \ref{fig:1068}). NGC 1068 is the closest Seyfert galaxy and is widely considered an archetype of this class \citep{Seyfert1943, Antonucci1985}. NGC 1068 has long been a high priority target for the astronomical community and has become one of the most well-studied active galaxies in the local universe. The nuclear core of NGC 1068 contains an AGN obscured by a dusty molecular torus \citep{Garcia-Burillo2014}. The central engine emits a radio jet \citep{Muxlow1996, Gallimore1996, Mutie2024} and drives a bi-conical radial outflow traced by line-emitting clouds \citep{Crenshaw2000, Cecil2002, Muller-Sanchez2011}. The central AGN is thought to have a widespread influence on the ionization properties of NGC 1068. A conical region of highly-ionized gas lies along the axis of the jet \citep{Pogge1988}, likely heated by nuclear emission from the AGN. A diffuse ionized medium seen in H$\alpha$ is distributed throughout the galactic disk along with ionized gas filaments traced by NII \citep{Bland-Hawthorn1991}. Far beyond the disk, faint H$\alpha$ knots/filaments are found aligned with both lobes of the optical ionization cone \citep{Veilleux2003}. A large star forming ring surrounds the galaxy at a radius of $\sim 11$ kpc \citep{Thilker2007}. In addition, ultra-diffuse features have been imaged around the galaxy within 45 kpc and are possibly indicative of a past minor-merger \citep{Tanaka2017-2}.  NGC 1068 has extensive multi-wavelength observations and has been observed by nearly every NASA astrophysics mission to date. Despite this pan-chromatic wealth of attention, much of the literature has focused on the active nucleus of NGC 1068, and far fewer papers (e.g., \cite{Thilker2007, Tanaka2017-2, Veilleux2003}) have been written about the large-scale extended structure surrounding the galaxy. 

In Section \ref{sec:observations}, we detail observations collected with \chas. We present the preliminary data reduction and analysis in Section \ref{sec:results}. Section \ref{sec:discussion} discusses the properties and origin of extended emission around NGC 1068 as well as important insights into the nature of interactions between the active galaxy and the larger circumgalactic medium (CGM). Key points from this publication are summarized in Section \ref{sec:summary}.

\section{Observations}
\label{sec:observations}
The observations presented in this paper were collected with the Circumgalactic H$\alpha$ Spectrograph (\chas{}) during its early commissioning phase at MDM Observatory. We describe the raw data and summarize the key image reduction techniques applied to create the data products presented in this work. They have been reduced using a preliminary version of the CH$\alpha$S data reduction pipeline, which is still under construction and will be described in full in a future publication.

\subsection{\chas{} Observations}
\label{subsec:chasobservations}

\chas{} is an integral field spectrograph collecting spatially resolved spectra or spectral imaging.  Each raw \chas{} image is a 3D spectral cube flattened onto a 2D detector. The field of view is sampled by a microlens array, resulting in a direct and stable mapping of the telescope focal plane. Each lenslet has a diameter of 2.55 arcsec and bins light over a 5 arcsec$^{2}$ field position into a smaller spot (an image of the telescope pupil).
Light from each lenslet spot is diffracted into a short ultra-narrowband spectrum on the detector. The spectrograph output is a Pointillist image comprised of $> 60,000$ spectra collected in a single exposure. The  CH$\alpha$S spectral images presented in this paper are not summed over wavelength. They are left as a 2-D grid of spectra where the dispersion extends along the x-axis. Dispersed stars appear as continua spectra, where the elongated width is equal to the bandpass of the narrowband filter. H$\alpha$ emission lines cover a much shorter spectral range; even if the line width is resolved, it will typically cover only a few resolution elements. Narrow linewidth Telluric emission from bright sky lines can be seen as a uniform background grid of spectrally unresolved spots from each lenslet. See \cite{Melso2022} for a full description of the instrument and data format. 

Spectral imaging of NGC 1068 is shown in Figure \ref{fig:1068}. 
The observational summary and instrument properties are given in Table \ref{table:obs}. This stack is a 6.4 hour integration collected over 3 nights ($3-5$ December 2021). The full integration is split into 64 frames, each with an exposure time of 360s. The exposure time is based on the instrument flexure \citep{Melso2022} and is selected to preserve the spectral imaging quality. The data was taken during exceedingly dark conditions; each exposure in the stack has a moon illumination of $< 2 \%$ and a moon separation of $> 125^{\circ}$. CH$\alpha$S samples the field of view with a spatial resolution of $2.8''$ and a resolving power of R$\sim 10,000$. Targets selected for CH$\alpha$S are chosen within a redshift range such that their H$\alpha$ emission is redshifted into the narrow filter bandpass. At the systemic velocity of NGC 1068, H$\alpha$ emission corresponds to a central wavlength of  $\rm 6588 \ \AA$. The observations presented in this paper use a combination of filters to achieve a bandpass of $\rm 20 \ \AA$ centered on a wavelength of $\rm 6582 \ \AA$. The FWHM filter response nearly rejects both lines in the redshifted NII doublet 
 ($\rm 6573 \ \AA$, $\rm 6608 \ \AA$) and excludes all but one bright sky emission line; the bright OH 6-1P$_{1e,1f}$(4.5) emission line (doublet) at 6578 $\rm \AA$ remains in-band \citep{Osterbrock1996}.

\subsection{\chas{} Data Reduction}
\label{subsec:reduction}
 
While work is still ongoing to refine and publish the CH$\alpha$S data reduction pipeline, this section briefly describes the procedures used to process data for this publication. Multiple exposures are aligned and stacked to build up long integration times. The drift/offset between exposures is calculated using the cross-correlation. This image registration is performed to sub-pixel precision, and the data is upsampled using a third-order spline. Cosmic rays are masked in each exposure and removed from the stack by looking for Laplacian/sharp edges following the prescription in \cite{VanDokkum2001}. Ghost features (including optical ghosts and cross-talk in the detector electronics) are rejected using multiple frames. They are often replicated in the four detector quadrants, mirrored/flipped in orientation, or appear to counter-drift with respect to the drift in the instrument field center. 

We generate a custom solution for the CH$\alpha$S sky-to-pixel mapping as a function of wavelength. In selecting a wavelength for the astrometric solution, we are unable to use prominent absorption features in the stellar spectra as the majority of stars are relatively featureless in the narrow \chas{} bandpass. The solution presented is centered on the H$\alpha$ emission line redshifted to the systemic velocity of NGC 1068. This provides the best average alignment between CH$\alpha$S emission features and narrowband imaging, although it does not capture position offsets of H$\alpha$ emission lines due to Doppler velocities with respect to systemic. All wavelength values are given in air and the Heliocentric velocity correction is noted in the header but not directly applied to the WCS. Simple Imaging Polynomial (SIP) Distortion coefficients are a work in progress and have not been applied \citep{Shupe2005}. 

A preliminary flux calibration has been applied to the data products presented in this work (See Figure \ref{fig:ribbondata}). This flux calibration is based on the R band magnitude of stars in the USNO UCAC5 catalog around NGC 1068 \citep{Zacharias2017}. In order to remove the sky lines, a local background subtraction is performed using a patch of clear sky. Sky background subtraction and flux calibration around the Ribbon feature is shown in Figure \ref{fig:ribbondata}. The sky image is aligned with the science image, subtracted, and saved for future reference. To assess the noise following background subtraction, we create a histogram of summed counts in a rolling rectangular aperture with a size of (w, h) = (5, 3) pixels corresponding to emission with a velocity width of $100 \rm \  km \ s^{-1}$. We aim for a distribution centered on zero with a standard deviation of less than a few counts.  The spatial response in the stacked images is handled by averaging and smoothing a collection flat field images collected in the corresponding filter combination. We divide the stacked, sky background subtracted image by the normalized flat field. Spectral extraction performed in (w, h)  = (5, 3) pixel rectangular apertures approximately centered on the detected emission from each lenslet. Each 2D extracted spectrum is collapsed to a 1D median spectrum and fit with a Gaussian to determine the peak centroid and standard deviation. We use the saved sky image as a reference grid, apply a similar spectral extraction, and calculate the offset between the detected emission and the restframe sky line at 6578 $\rm \AA$. This offset is converted to a measurement of the gas velocity using the systemic velocity of the galaxy and the heliocentric velocity correction. The velocity and dispersion solution is shown in Figure \ref{fig:ribbonvelocity}. 

\begin{deluxetable}{lll}
\tablecaption{NGC 1068 ribbon observations}
\label{table:obs}
\tablehead{
\colhead{Parameter} & \colhead{Value} & \colhead{Comment}
}
\startdata
\textbf{Observation Summary} &  &  \\
Right Ascension ($\alpha$) & 02:42:54 & FOV Center\\
Declination ($\delta$) & -00:00:16 & FOV Center \\
Distance & 14.4 Mpc & \cite{Tully1988}\\
Systemic (V$_{\rm Hel}$) & 1137 $\pm$ 3 km s$^{-1}$ & \cite{Allison2014}\\
Exposure Time & 360 s & per frame\\
Total Integration  & 6.4 hrs  & 64 frames \\ 
Moon Illumination & $< 2\%$ & \\
Moon Separation & $> 125^{\circ}$ & \\
Dates &  $3-5$ Dec 2021 & \\
Heliocentric Correction & $-16.454 \ \rm km \ s^{-1}$ & \\
\hline
\textbf{Instrument Settings} &  &  \\
Field of View & $10' \times 10'$ & \\
Spatial Resolution & $2.8 ''$ & \\
Spectral Resolution $(\Delta \lambda)$ & $0.67 \rm \ \AA$ & \\
Central Wavelength $(\lambda)$ &  $\rm 6582 \rm \ \AA$ &\\ 
Bandpass (FWHM) & $\rm 20 \ \AA$ & \\
\hline 
\textbf{Ribbon Properties} &  &  \\
Right Ascension ($\alpha$) & 02:42:59.24 & J2000 \\
Declination ($\delta$) & +00:00:47.84 & J2000 \\
Projected Radius & 20.4 kpc & $1'' = 72$ pc \\
H$\alpha$ Flux & $[4-16]$ R \tablenotemark{a} & See Figure \ref{fig:1068}\\
\enddata
\tablenotetext{a}{1 R = $5.66 \times 10 ^{-18}$ ergs s$^{-1}$ cm$^{-2}$ arcsec$^{-2}$} 
\end{deluxetable}

\subsection{Ancillary Data}
We compare with {\it GALEX} NUV and FUV images of NGC 1068. All the The {\it GALEX} data presented in this article were obtained from the Mikulski Archive for Space Telescopes (MAST) at the Space Telescope Science Institute. The specific observations analyzed can be accessed via \dataset[10.17909/abmt-8813] {http://dx.doi.org/10.17909/abmt-8813}. UV background subtraction is performed using the {\it GALEX} pipeline skybkg files. We correct all measurements using the {\it GALEX} response curve in the NUV ($\lambda_{eff} = 2267 \AA$) and FUV ($\lambda_{eff} = 1516$) band \citep{Martin2005, Morrissey2005, Morrissey2007}. In order to confirm extended H$\alpha$ features, we compare with ancillary narrowband H$\alpha$ imaging \citep{Knapen2004, Veilleux2003, Brown2014} around NGC 1068.

We compare NGC 1068 with other galaxies hosting extended emission line regions. This work incorporates H$\alpha$ data around NGC 5128 (Centaurus A) downloaded from the NOIRlab Archive (Ian Evans, private communication). These narrow-band (8 nm) images were obtained using the Mosaic 2 imager on the CTIO 4m telescope. The stacked image data product has gone through the NOAO Mosaic Pipeline. Photometric calibration was applied using the pipeline zero-point magnitude for this stack. 

\section{Results}
\label{sec:results}

\begin{figure*}
\centering
\includegraphics[scale=1]{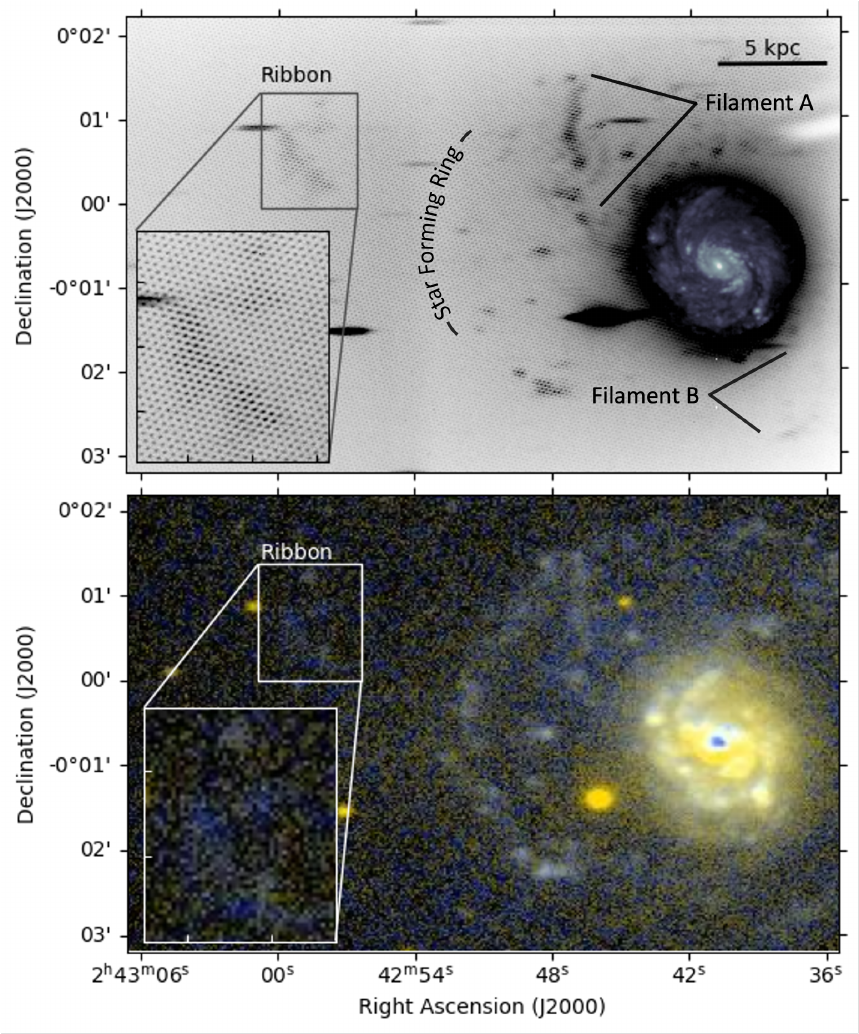}
\caption{\small \em{[Top] CH$\alpha$S spectral imaging of NGC 106. The CH$\alpha$S data is comprised of a grid of spectra; dispersed stars appear as elongated continua spectra while narrow linewidth emission (e.g., bright sky lines, H$\alpha$) results in short spectra comprised of only a few resolution elements.} An SDSS u-band image of the NGC 1068 galactic disk is overlaid for reference \citep{Abazajian2009}. The CH$\alpha$S data presented here is not sky subtracted; the background is dominated by the bright OH 6-1P$_{1e,1f}$(4.5) emission line (doublet) at 6578 $\rm \AA$. We label the Ribbon feature (with zoom-in) and other previously known structure including filament A, filament B and the star forming ring. [Bottom] {\it GALEX} FUV (blue) and NUV (yellow) composite image with WCS applied. We label the Ribbon feature (with zoom-in). \label{fig:1068}}
\end{figure*}

We present new wide-field spectral imaging of NGC 1068 with the Circumgalactic H$\alpha$ Spectrograph in Figure \ref{fig:1068}. This image is not background subtracted; the Telluric OH 6-1P$_{1e,1f}$(4.5) emission line can be seen as a uniform background grid of spots from each lenslet.

\subsection{Extended Gas Morphology}
The ionized gas distribution around NGC 1068 includes features ranging in size from $20''-150''$ with variations occurring on the order of a few arcseconds. These variations are easily seen in the CH$\alpha$S data, which probes clumping down to spatial scales of $\sim 200 \ \rm  pc$ $(2.8'')$. In the CH$\alpha$S data we recover the extended filamentary structure discussed in \cite{Veilleux2003}, labeled as Filament A and Filament B. This biconical filamentary structure is roughly constrained to position angles aligned with the axis of the central jet (PA $\sim 30^{\circ}$) \citep{Gallimore2001, Garcia-Burillo2014}, and it lies within the extrapolated optical ionization cone emanating from the AGN \citep{Pogge1988}. The tip of these filaments overlap in projection with a large star forming ring seen prominently in {\it GALEX} FUV imaging at a galactocentric radius of $\sim 11$ kpc \citep{Thilker2007}. We detect bright clumps/HII regions coincident with this ring along the labeled arc in Figure \ref{fig:1068}, as seen in previous H$\alpha$ imaging \citep{Knapen2004, Brown2014}. We also detect previously unreported extended emission located beyond these filament/ring structures towards the edge of the CH$\alpha$S field of view. This new emission is labeled \textit{Ribbon} in Figure \ref{fig:1068} and it lies just outside the deep, narrowband imaging presented in Veilleux et al. 2003 with a sky coordinate of approximately ($\alpha, \delta$) = (02:42:59.24, +00:00:44.43). This places it well into the galactic halo at an angular distance of $292 ''$ from the galactic center ($\alpha$, $\delta$) = (02:42:40.77, 00:00:47.84) and a projected physical radius of 20.4 kpc given a standard distance d = 14.4 Mpc \citep{Bland-Hawthorn1997}. On closer inspection, the Ribbon is visible at low signal-to-noise in both the {\it GALEX} NUV and FUV images as well as in narrowband H$\alpha$ imaging \citep{Knapen2004, Brown2014}. It spans a position angle of approximately $65^{\circ} - 80^{\circ}$, in relative proximity to the jet axis but near the edge of the extrapolated optical ionization cone measured on small scales \citep{Pogge1988} with an opening angle of $\sim 40^{\circ}$ and a position angle of $\sim 32^{\circ}$. However, some models for the biconical outflow region quote a wider $\sim 80^{\circ}$ opening angle that could intersect the Ribbon emission \citep{Crenshaw2000, Das2006, Poncelet2008}. The morphology of the Ribbon feature around NGC 1068 is fairly similar to the weather ribbon of FUV and H$\alpha$ emission detected far into the halo of NGC 5128 around Centaurus A \citep{Morganti1991, Neff2015}; the twisted Ribbon of linear structure ends in a v-shaped filament tens of kpc from the galactic disk.

\subsection{Extended Gas Kinematics}
\label{sec:kinematics}

\begin{figure*}
\centering
\includegraphics[scale=0.3]{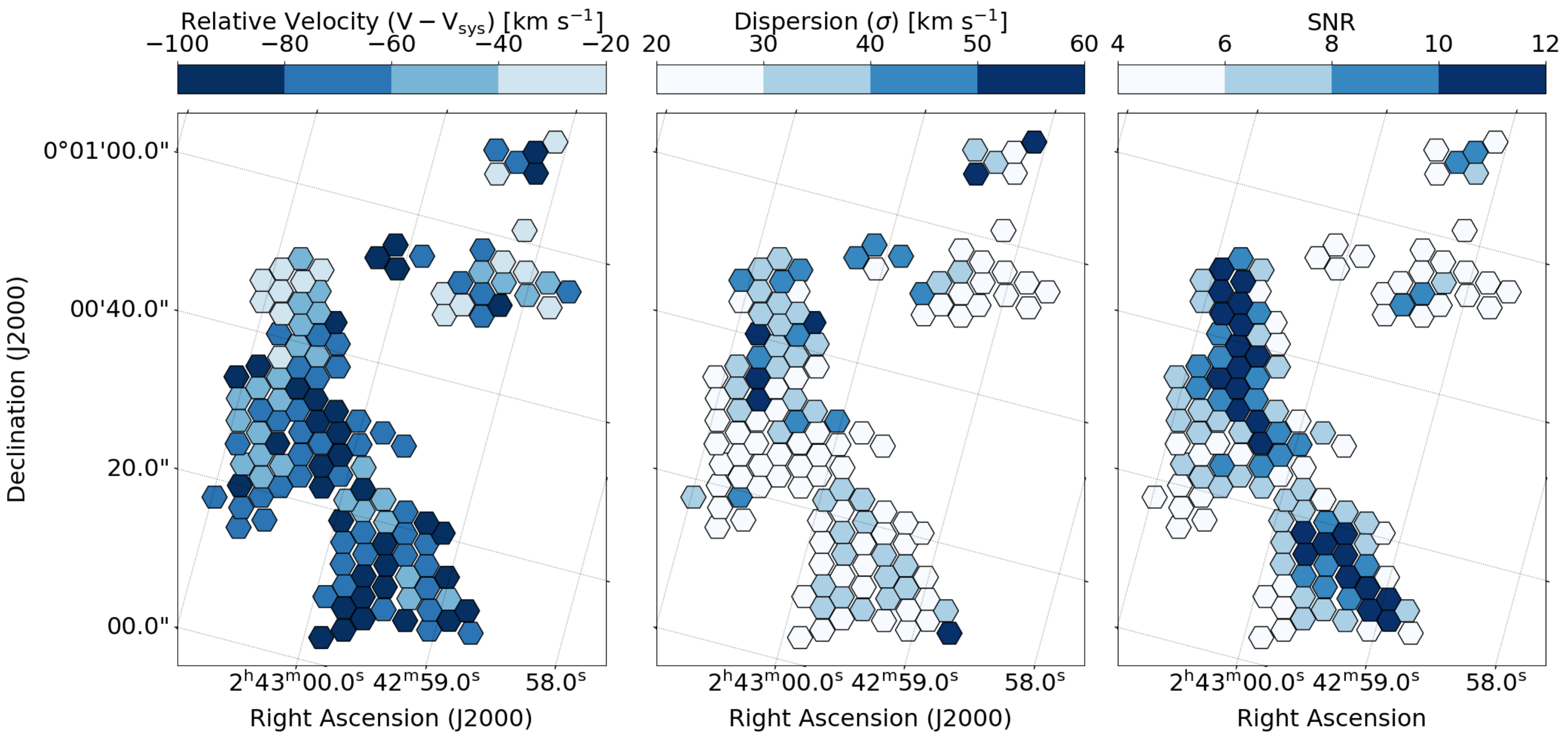}
\caption{\small \em{CH$\alpha$S spectral fitting for the Ribbon. All panels have been rotated ($\sim 15^{\circ}$) to align the hexagonally packed spectra along the Cartesian y-axis. The transformed coordinates can be compared directly with Figure \ref{fig:1068}. (left) Relative velocity with respect to the systemic velocity of NGC 1068 $(\rm V_{sys} = 1137 \ km \ s^{-1})$ from Gaussian fit of each lenslet spectrum. (middle) Velocity dispersion from Gaussian fit to each lenslet spectrum. (right) Spectra analyzed have a SNR $\geq 5$ in a rectangular aperture (w, h) = (5, 3) pixels. This aperture sizing corresponds to emission with a velocity width of $\sim 100 \rm \  km \ s^{-1}$}. \label{fig:ribbonvelocity}}
\end{figure*}

Sky subtracted stacks were used to create velocity and dispersion maps for the extended ionized gas features as shown in Figure \ref{fig:ribbonvelocity} and Figure \ref{fig:combdata}. The spectral extraction is described in Section \ref{subsec:reduction}.
Relative velocity values are referenced with respect to the galaxy systemic (1137 km s$^{-1}$) \citep{Allison2014} with the Heliocentric correction in Table \ref{table:obs}. All of the extended ionized gas along the east side of the galaxy imaged with CH$\alpha$S is blue-shifted in agreement with the HI rotation map \citep{Brinks1997}, but with a small velocity gradient across the Ribbon on the order of 40 km s$^{-1}$.  Average values from the CH$\alpha$S velocity calibration have been confirmed using follow-up long slit spectra collected along the Ribbon and Filament A using the Ohio State Multi-Object Spectrograph (OSMOS) installed on the MDM Hiltner 2.4-m telescope \citep{Martini2011}. The Ribbon gas exhibits a slight velocity and dispersion gradient, with lower velocity and higher dispersion at the NE tip of the Ribbon. However, the dispersion gradient is comparable to the uncertainty on the dispersion measurement. The velocity offset and line broadening for Filament A is consistent with long-slit spectra presented in \cite{Veilleux2003}. Velocities in the star forming ring are similar to the velocity of the Ribbon gas but slightly higher where the ring intersects Filament A.
 
\subsection{Additional Properties}
\label{sec:addprop}
\begin{figure*}
\centering
\includegraphics[scale=0.3]{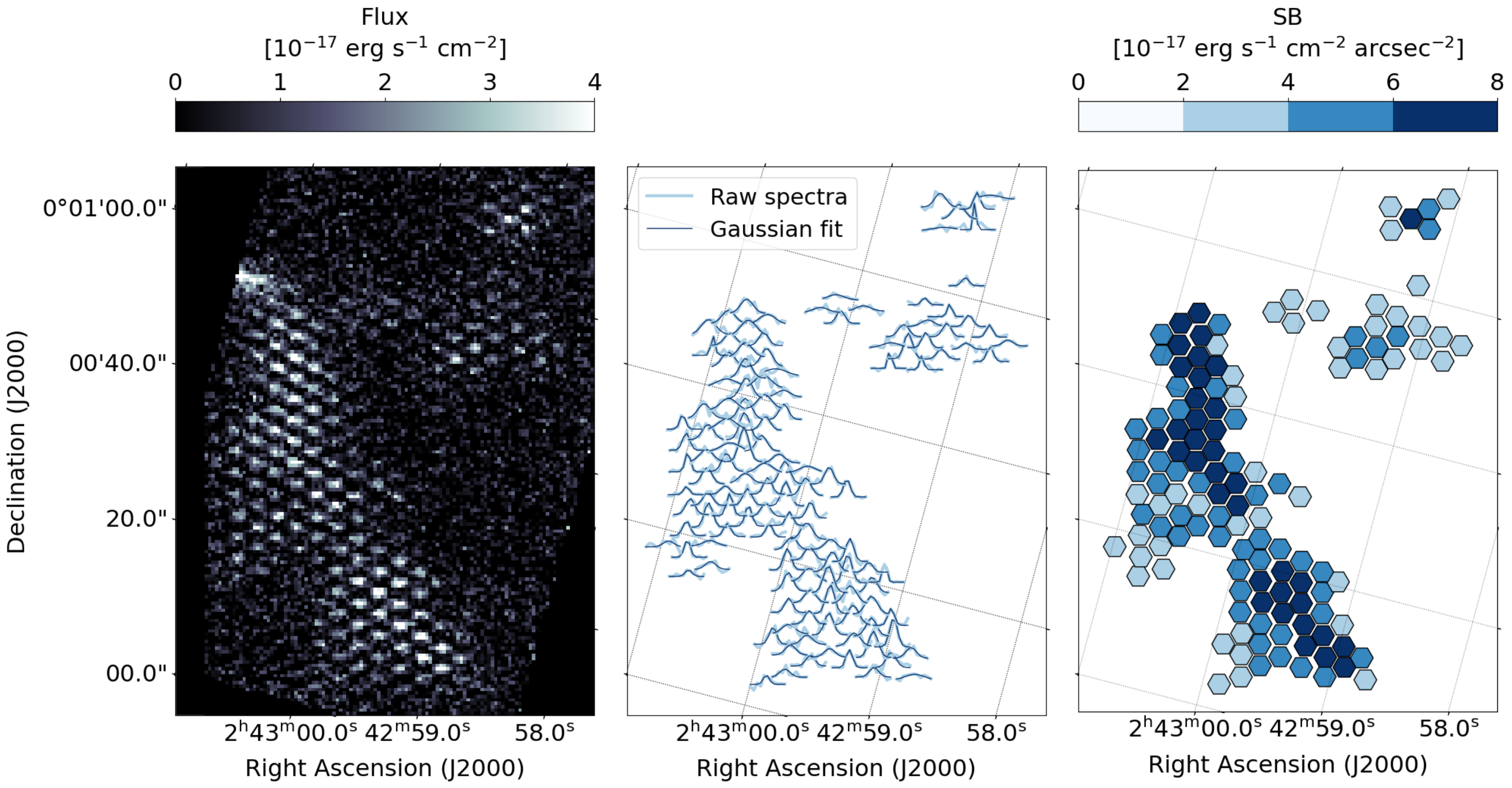}
\caption{\small \em{CH$\alpha$S spectral extraction and flux calibration for the Ribbon. All panels have been rotated ($\sim 15^{\circ}$) to align the hexagonally packed spectra along the Cartesian y-axis. The transformed coordinates can be compared directly with Figure \ref{fig:1068}. (left) Flux in each spatial-spectral pixel of the sky-background subtracted Ribbon data. (middle) Raw spectra extracted from each lenslet and the Gaussian fit. The spectral amplitude has been scaled to reduce overlap in this plot; however, the relative amplitude between spectra remains unchanged. (right) Surface brightness in each lenslet (2.55 arcsecond diameter, 5 arcsec$^{2}$ solid angle) extracted with aperture photometry. We sum the flux across each spectrum aperture, integrating over the linewidth, and then divide by the lenslet solid angle. \label{fig:ribbondata}}}
\end{figure*}

\begin{figure*}
\centering
\includegraphics[scale=0.3]{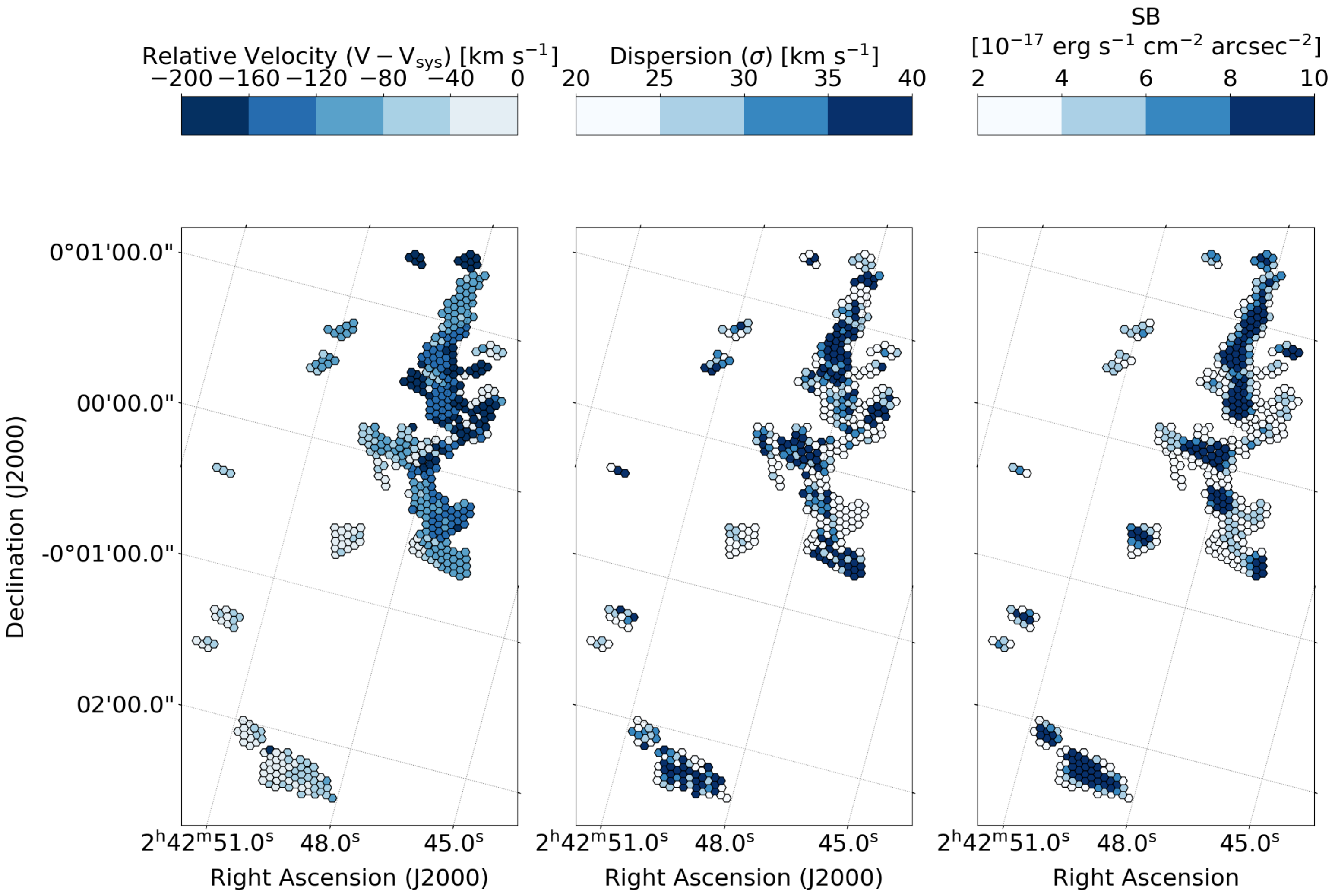}
\caption{\small \em{CH$\alpha$S spectral extraction and flux calibration for Filament A and the Ring. All panels have been rotated ($\sim 15^{\circ}$) to align the hexagonally packed spectra along the Cartesian y-axis. The transformed coordinates can be compared directly with Figure \ref{fig:1068}. (left) Relative velocity with respect to the systemic velocity of NGC 1068 $(\rm V_{sys} = 1137 \ km \ s^{-1})$ from Gaussian fit of each lenslet spectrum. (middle) Velocity dispersion from Gaussian fit to each lenslet spectrum. (right) Surface brightness in each lenslet (2.55 arcsec diameter, 5 arcsec$^{2}$ solid angle) extracted with aperture photometry. \label{fig:combdata}}}
\end{figure*}

We compare the properties of prominent extended features in our spectral imaging including Filament A, the star-forming ring, and the new Ribbon emission (Filament B is too close to the edge of our field of view). Measured properties are presented in Table \ref{table:measured} and Table \ref{table:galex_measured} while derived properties are presented in Table \ref{table:derived}. Properties are measured or derived in all lenslet apertures with detected H$\alpha$ emission in each feature. The apertures used are shown in Figure \ref{fig:ribbondata} and Figure \ref{fig:combdata}. The values measured or derived in each feature are quoted as a mean and (1$\sigma$) scatter about the mean representative of the range in values.
Measured properties from the CH$\alpha$S data include the surface brightness $\Phi_{H\alpha}$, angular scale $\Omega$, physical scale, corresponding H$\alpha$ Luminosity L(H$\alpha$), and emission measure. The angular scale $\Omega$ is the summed area of all the lenslet apertures with detected H$\alpha$ emission in each feature (see Figure \ref{fig:ribbondata} and Figure \ref{fig:combdata}). The NUV and FUV Luminosity were measured using {\it GALEX} images of NGC 1068. Derived properties include the star formation rate $\rm SFR(H\alpha)$, star formation rate surface density $\Sigma_{SFR}$, gas surface density $\Sigma_{gas}$, and HI column density $N_{HI}$. The measured H$\alpha$ luminosity is converted to an implied star formation rate following the prescription in \cite{Calzetti2013}. We derive the corresponding gas surface density along the Kennicutt-Schmidt relation \citep{Kennicutt1998}. 

One peculiarity of these measurements is that the ratio of H$\alpha$/UV emission is increasing with radius from the galactic center (See Figure \ref{fig:ratios}). While H$\alpha$ emission is certainly found at large radii and coincident with extended UV (XUV) disks (e.g., \cite{Lelievre2000, Werk2010, Goddard2011}), the H$\alpha$ radial emission profile typically falls off beyond the optical edge of the galactic disk \citep{Kennicutt1989, Martin2001, Goddard2010}. Some of the extended features around NGC 1068 (such as Filament A) have UV structure that is optically well-matched, resulting in \cite{Thilker2007} classifying NGC 1068 as a borderline Type 1 XUV disk similar to NGC 4736 and NGC 1291. In contrast, the extended star forming ring around NGC 1068 appears to be a more classical example of a UV bright/optically faint XUV disk feature. The discovery of the Ribbon gas extends the trend of enhanced H$\alpha$ emission at large radii around NGC 1068 out to $\sim 20$ kpc. We discuss the possible implications of this extended H$\alpha$ excess in the Ribbon gas in Section \ref{subsec:excitation}. 

In the top panels of Figure \ref{fig:ratios} we plot the ratio of H$\alpha$ line flux to the {\it GALEX} NUV ($\lambda_{eff} = 2267 \AA$, FWHM = $616 \AA$) and FUV ($\lambda_{eff} = 1516$, FWHM =$269 \AA$) \citep{Martin2005, Morrissey2005, Morrissey2007} continuum flux ($f_{H\alpha}/f_{UV})$ in units of $\rm \AA$ as a function of H$\alpha$ surface brightness. Values for NGC 1068 have been corrected for Galactic extinction following the prescription in \cite{Cardelli1989} using E(B-V) = 0.034 measured from Milky Way foreground reddening maps 
\citep{Schlegel1998}. The data have not been corrected for intrinsic extinction in the halo of NGC 1068 itself. UV background subtraction was performed using the {\it GALEX} pipeline skybkg files. 

In all panels of Figure \ref{fig:ratios} we compare the Ribbon emission around NGC 1068 with the analogous weather ribbon of FUV and H$\alpha$ emission detected far into the halo of NGC 5128 around Centaurus A (blue contours). H$\alpha$ data for the weather ribbon was downloaded from the NOIRlab Archive (Ian Evans, private communication). These narrow-band (8 nm) images were obtained using the Mosaic 2 imager on the CTIO 4m telescope. The stacked image data product has gone through the NOAO Mosaic Pipeline. Photometric calibration was applied using the pipeline zero-point magnitude for this stack. Values for emission around Centaurus A have been corrected for Galactic extinction following the prescription in \cite{Cardelli1989} using E(B-V) = 0.10 measured from Milky Way foreground reddening maps \citep{Schlegel1998}. We also compare with the H$\alpha$/UV flux ratios measured in the halos of NGC 283 and M82 \citep{Hoopes2005} converted to units of $\rm \AA$ (plotted as grey open circles in Figure \ref{fig:ratios}). Extended UV emission surrounding NGC 283 and M82 is similar in morphology to the extended H$\alpha$ emission. These values have also been corrected for Galactic extinction but not intrinsic extinction (see \cite{Hoopes2005}).

\subsection{Ionization Modeling}
\label{sec:modeling}
\begin{figure*}
\centering
\includegraphics[width=0.8\textwidth]{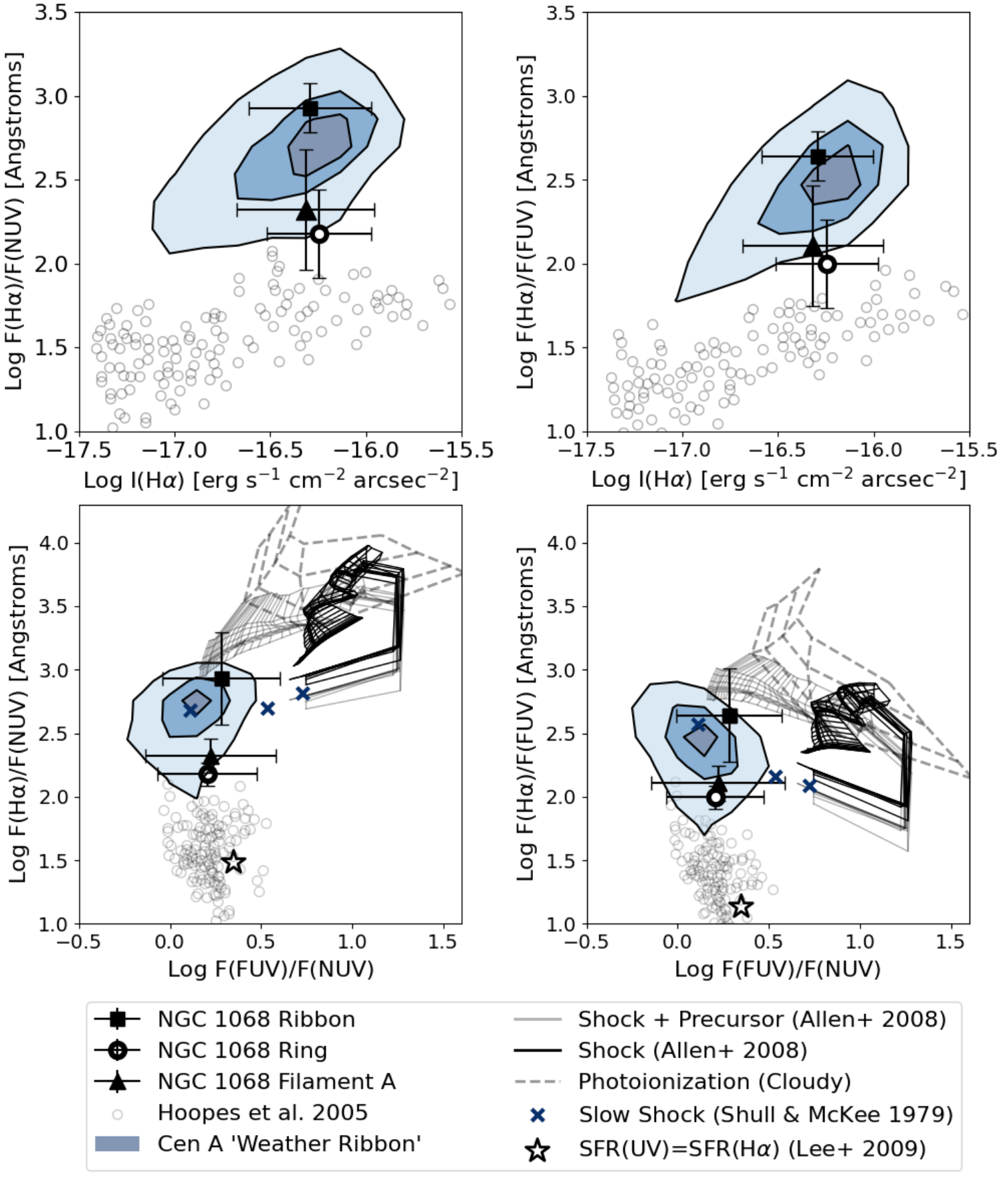}
\caption{\small \em{UV-optical diagnostic plots. Measurements for the Ribbon, Filament A, and the Ring around NGC 1068 are shown as a square, triangle and ring respectively. Error bars represent the variation across each feature. We compare with existing measurements in the halos of NGC 253 and M82 \citep{Hoopes2005} denoted by small grey open circles. We also compare with the of FUV and H$\alpha$ emission detected far into the halo of NGC 5128 around Centaurus A \citep{Neff2015}. Contours representing the Centaurus A weather ribbon measurement $[1\sigma, 1.5\sigma, 2\sigma]$ are shown in blue. All measured values have been corrected for foreground Galactic reddening (see Section \ref{sec:addprop}). (top left) Ratio of H$\alpha$ line flux to {\it GALEX} NUV flux density in units of $\AA$ as a function of H$\alpha$ surface brightness. (top right) Ratio of H$\alpha$ line flux to {\it GALEX} FUV flux density in units of $\AA$ as a function of H$\alpha$ surface brightness. (bottom left) Diagnostic plot showing the ratio of H$\alpha$ line flux to {\it GALEX} NUV flux density in units of $\AA$ vs. the unitless {\it GALEX} FUV to {\it GALEX} NUV flux density ratio. (bottom right)} Diagnostic plot showing the ratio of H$\alpha$ line flux to {\it GALEX} FUV flux density in units of $\AA$ vs. the unitless {\it GALEX} FUV to {\it GALEX} NUV flux density ratio. We compare with models for photoionization, shocks, and shocks with photoionized precursors. Model details are discussed in Section \ref{sec:modeling} \label{fig:ratios}}
\end{figure*}

In this section we explore the possibility that the FUV and NUV light from the Ribbon may be dominated either by UV emission lines, or by non-ionizing UV stellar continuum from young stars. Many EELRs observed with GALEX exhibit a combination of UV continuum emission and line emission such as C IV, He II and C III] (e.g., \cite{Keel2012}). The bottom panels in Figure \ref{fig:ratios} are UV-optical line ratio diagrams of H$\alpha$/NUV and H$\alpha$/FUV vs FUV/NUV. UV and UV-optical line ratio diagrams are valuable diagnostics for discriminating between shock and photoionization excitation models \citep{Allen1998} and emission dominated by recent star-formation. We overplot models for photoionization, shocks, and shocks with ionized precursors assuming solar metallicity (1 Z$_{\odot}$) gas with a (pre-shock) density of 1 cm$^{-3}$. We have performed a similar analysis at sub-solar metallicity and find that emission from the Ribbon gas is not as consistent with photoionization or shock models at 0.5 $\rm Z_{\odot}$. In order to compare with the continuum ratios measured around NGC 1068, we select bright emission lines in the {\it GALEX} NUV (C III]($\lambda 1909$), C II] ($\lambda 2326$)) and FUV (C IV ($\lambda 1549$), He II ($\lambda 1640$)) bandpass and sum their emission intensities weighted by the {\it GALEX} response as a function of wavelength. To convert to units of Angstroms, we divide by the continuum bandpass. 

The photoionization models are equilibrium models run in CLOUDY \citep{Ferland2013} using plane-parallel geometry to represent a cloud far from the ionizing sources in the galactic disk. The incident radiation is set to a simple power-law continuum using the default energy limits and a variable slope. The model grid spans a range of spectral index $\alpha = [-2, -1.7, -1.4, -1.2]$ and ionization parameter $ \rm log \ U = [-4, -3, -2, -1]$. These models assume solar abundance 1Z$_{\odot}$ with a hydrogen density of 1 cm$^{-3}$. The run is stopped when the electron fraction $\rm n(H^{+})/n(H)$ drops below 0.01. The results are very similar to the solar abundance, no-dust photoionization model from \cite{Groves2004a} derived using the photoionization and shock code MAPPINGS III 
\citep{Kewley2001, Groves2004a, Sutherland1993}. 

We adopt fast shock models from \citep{Allen2008}. These are equilibrium models run using MAPPINGS III. The models include gas cooling behind the shock front, which can produce strong ionizing radiation resulting in a photoionized precursor. This model covers a range in shock velocity from $100 - 1000 \rm \ km \ s^{-1}$ and varying magnetic field strength including $B = [$1E-4$, 0.5, 1.0, 2.0, 3.23, 4.0, 5.0, 10.0] \ \mu G$. We plot the solar abundance model with a pre-shock density of $\rm 1 \ cm^{-3}$. We also compare with slow shock models from \cite{Shull1979} covering range in velocity of $40-130$ km s$^{-1}$. These models use a pre-shock density of 10 cm$^{-3}$, a magnetic field strength of $1 \mu G$, and cosmic metal abundances. 

In addition to comparing with the above ionization models, we also equate the SFR inferred from the UV non-ionizing continuum and H$\alpha$ nebular emission. We equate the SFR in $\rm M_{\odot} \ yr^{-1}$ using the equations from \cite{Lee2009} $\rm SFR(H\alpha) = 7.9 \times 10^{-42} L(H\alpha) \ [erg \ s^{-1}]$ and $\rm SFR(UV) = 1.4 \times 10^{-28} L_{\nu}(UV) \ [erg \ s^{-1} \ Hz^{-1}]$. The UV luminosity values are converted to units of $\rm ergs \ s^{-1} \ \AA^{-1}$ for the {\it GALEX} NUV ($\lambda_{eff} = 2267 \AA$) and FUV ($\lambda_{eff} = 1516$) band. The star-shaped marker in Figure \ref{fig:ratios} denotes UV-optical ratios that equate the SFR derived from H$\alpha$ and UV, implying excitation by in-situ star formation.

\begin{deluxetable*}{cccccc}
\tablecaption{CH$\alpha$S Measured properties}
\label{table:measured}
\tablehead{\colhead{Feature} & \colhead{$\rm SB(H\alpha$)} & \colhead{$\Omega$} & \colhead{Scale} & \colhead{L(H$\alpha$)} & 
\colhead{EM} \\ & [erg s$^{-1}$ cm$^{-2}$ arcsec$^{-2}$] & [arcsec$^{2}$] & [kpc$^{2}$] & [$\times 10^{38}$ erg s$^{-1}$] & [$\rm cm^{-6} \ pc$]}
\startdata
Filament A & $ 6.69 \pm 5.33 \times 10^{-17} $ & 1790 & 8.72 & $ 29.7 \pm 23.7 $ & $33.11 \pm 26.34$ \\ 
Ribbon & $ 5.44 \pm 1.80 \times 10^{-17} $ & 640 & 3.12 & $ 8.63 \pm 2.86 $ & $26.89 \pm 8.91$\\
SF Ring & $ 7.71 \pm 5.17 \times 10^{-17} $ & 540 & 2.63 & $ 10.3 \pm 6.9$ & $38.13 \pm 25.58$\\ 
\enddata
\tablenotetext{a}{1 R $\rm =  2.8 \ cm^{-6} \ pc$ (EM, Case B)} 
\end{deluxetable*}

\begin{deluxetable*}{ccccc}
\tablecaption{GALEX Measured properties}
\label{table:galex_measured}
\tablehead{\colhead{Feature} & \colhead{SB(NUV)} & \colhead{SB(FUV)} & \colhead{$f_{H\alpha}/f(NUV)$} & \colhead{$f_{H\alpha}/f(FUV)$} \\ & [$\rm erg \ s^{-1} \ cm^{-2} \ arcsec^{-2} \ \AA$] & [$\rm erg \ s^{-1} \ cm^{-2} \ arcsec^{-2} \ \AA$] & [$\rm \AA$] & [$\rm \AA$]}
\startdata
Filament A & $2.73 \pm 1.78 \times 10^{-19}$ & $4.52 \pm 2.91 \times 10^{-19}$ & 290.0 & 175.0\\
Ribbon & $6.52 \pm 4.11 \times 10^{-20}$ & $1.42 \pm 0.95 \times 10^{-19}$ & 1388.0 & 478.0\\
Ring & $5.39 \pm 3.90 \times 10^{-19}$ & $8.80 \pm 6.97 \times 10^{-19}$ & 175.0 & 111.0\\
\enddata
\end{deluxetable*}

\vspace{-1.5cm}
\section{Discussion}
\label{sec:discussion}

The discovery of the extended ionized Ribbon feature raises important questions about the gas origin and excitation mechanisms. Here we discuss conclusions that can be drawn from the data in-hand, and place the Ribbon gas in the context of other extended emission line regions in the local universe. Follow-up observations of the neutral atomic and
molecular gas in the Ribbon feature will be important for understanding the origin of the gas. Additionally, future measurements of optical line ratios in the Ribbon will help distinguish between excitation mechanisms and characterize the anisotropic radiation field around NGC 1068 on large scales.

\subsection{The Origin of the Ribbon Gas}
\label{sec:origin}

The Ribbon gas is not associated with any galactic structure or known tidal features in the halo of NGC 1068. Despite the presence of high velocity outflows in NGC 1068 \citep{Cecil2002, Groves2004b}, it is unlikely that this gas was ejected from the disk/AGN as galactic feedback and survived the transit to 20 kpc. Clouds swept up and carried outward by a fast, hot galactic wind are easily destroyed. Assuming the Ribbon gas is moving away from the galaxy at a speed similar to the line-of-sight velocity ($\sim 80$ km s$^{-1}$), the Ribbon gas would take $\sim 250$ Myrs to traverse 20 kpc, much longer than the timescale required for the cloud to evaporate or end up shredded by fluid instabilities \citep{Klein1994}. 

One of the primary origin scenarios for the Ribbon gas is that NGC 1068 is surrounded by a distribution of cold gas and the Ribbon feature was ionized in situ. An extended distribution of cold gas is prevalent in interacting and merging galaxies. In NGC 1068, an extended star forming ring \citep{Thilker2007} and ultra-diffuse features within 45 kpc may be indicative of a past minor-merger \citep{Tanaka2017-2}. In Section \ref{sec:addprop} we use the observed H$\alpha$ luminosity in the Ribbon to estimate the HI column density around NGC 1068, which ranges from $10^{20}-10^{21} \rm \ cm^{-2}$ (See Table \ref{table:derived}). This column density is on par with expectations around other nearby galaxies. For example, the HI distribution around NGC 5128 was mapped to a limiting column density of $\rm 10^{20} \ cm^{-2}$, and the 17 kpc HI cloud has a peak column density of $1.7 \times 10^{21} \rm \ cm^{-2}$ \citep{Schiminovich1994, Oosterloo2005}. Near the Ribbon feature, localized peaks of the HI distribution are likely to be small-scale over-densities a factor of a few higher in column density. While we assume that the majority of the gas is neutral, a high ionization fraction could skew this estimate to lower column densities. Conversely, a low star formation efficiency leading to a long gas consumption timescale would imply a much more massive HI distribution.

\subsection{Excitation of the Ionized Gas}
\label{subsec:excitation}

The ionized gas Ribbon in NGC 1068 likely needs an ongoing excitation source, as the radiative cooling timescale $\rm t_{cool} = kT_{c}/n\Lambda(T_{c})$ is $< 10,000$ years assuming a density of $> 0.05$ cm$^{-3}$ (Section \ref{sec:origin}) and a cooling rate of $\rm \Lambda(T_{c}) = 10^{-22} \ ergs \ cm^{3} \ s^{-1}$ at $T_{c} = 10^{4}$. For higher density features, on the order of 10 cm$^{-3}$, the radiative cooling timescale is $< 50$ years. We discuss a number of models for exciting the extended emission line regions around NGC 1068 including photoionization, collisional/shock ionization, and in-situ star formation. The UV-optical ratios measured in NGC 1068 are consistent with extended emission line regions in halos of NGC 5128, NGC 253 and M82 \citep{Hoopes2005, Neff2015}. The UV flux from these extended features is generally too high to be consistent with emission dominated by fast shock-heating or photoionization and too low to be dominated by in-situ star formation alone. It is likely that the extended emission line regions around NGC 1068 are excited by a combination of processes. 

\subsubsection{Photoionization}
Galaxies with heightened star formation or active galactic nuclei leak an excess of photoionizing flux into their halos, ionizing gas far beyond the galactic disk. We estimate the HI photoionization rate required to produce the Ribbon emission using the observed H$\alpha$ surface brightness \citep{Donahue1995, Shull1999, Weymann2001, Adams2011, Fumagalli2017}. The photoionization rate ($\Gamma$) is given by Equation \ref{eq:prate}, where h is Planck's constant, $a_{\nu}$ is the cross section at the Lyman Limit ($6.3\times10^{-18} \ \rm cm^{2}$), $\rm J_{\nu} = J_{0}(\nu_{0}/\nu)^{\beta}$ is power-law spectrum of the ionizing photons, and $\beta$ is the power-law slope usually modeled as 1.8.

\begin{equation}
\label{eq:prate}
    \Gamma = \frac{4\pi a_{v}}{h} \frac{J_{0}}{\beta + 3}
\end{equation}

In Equation \ref{eq:J}, $\rm J_{0}$ (in ergs s$^{-1}$ cm$^{-2}$ sr$^{-1}$ Hz$^{-1}$) is related to the H$\alpha$ surface brightness $\Phi_{H\alpha}$ (in photons s$^{-1}$ cm$^{-2}$ sr$^{-1}$) and the ionizing specific intensity $\phi_{0}$  \footnote{$\phi$ is sometimes defined as the ionizing specific flux. Note the units here are intensity. } (in ph s$^{-1}$ cm$^{-2}$ sr$^{-1}$). 

\begin{equation}
\label{eq:J}
    J_{0} = \frac{h \Phi_{H\alpha}}{f_{H\alpha}} f_{\beta} f_{g} = h\phi_{0}f_{\beta}
\end{equation}
For case B recombination at a temperature of T $\sim 10^{4}$ K the fraction of incident ionizing photons converted to H$\alpha$ photons is $\sim 45\%$ ($f_{H\alpha}$ = 0.45). This calculation uses a spherical geometry $f_{g}$ = 1 and spectral shape parameter $f_{\beta} = \beta/(1-4^{-\beta})$ where we assume $\beta$ = 1.8. For an H$\alpha$ surface brightness of $\Phi_{H\alpha} = \rm 1.2 \times 10^{6} \ ph \ s^{-1} \ cm^{-2} \ sr^{-1} \ (15 \ R)$ in the brightest regions of the Ribbon gas, we calculate  $J_{0} = 3.45\times 10^{-20}$ ergs s$^{-1}$ cm$^{-2}$ sr$^{-1}$ Hz$^{-1}$, $\phi_{0} = 2.65\times 10^{6}$ ph s$^{-1}$ cm$^{-2}$ sr$^{-1}$, and $\Gamma = 8.57\times 10^{-11}$ s$^{-1}$. This photoionization rate is three orders of magnitude greater than current constraints on the extragalactic UV background (EUVB) photoionization rate ($\Gamma \lesssim 2-8 \times 10^{-14} \rm \ s^{-1}$) at low-redshift \citep{Adams2011, Madau2015, Fumagalli2017, Caruso2019}. If photoionized, a much stronger local source of ionizing radiation must be exciting the Ribbon gas around NGC 1068. One interpretation is that a beamed flux of photoionizing radiation from the nucleus of NGC 1068 excites gas within a bi-conical region of influence. This process has been suggested for the two inner filamentary features around NGC 1068 (Filament A and Filament B), which are better aligned with the current axis of the inner jet \citep{Veilleux2003}. The Ribbon gas, while not well aligned with the jet axis, may still lie within the ionization bicone emanating from the AGN \citep{Crenshaw2000, Das2006}. However, the UV-optical diagnostic ratios in Figure \ref{fig:ratios} for both the inner filamentary features as well as the outer Ribbon gas are not simply explained by a hard photoionizing radiation field (see Cloudy model in Figure \ref{fig:ratios}).

\subsubsection{Star Formation}
These faint outlying HII regions coincident with extended UV flux may be attributed to knots of newly formed massive stars \citep{Werk2010, Goddard2011}. Massive stars embedded in warm ionized gas produce a combination of H$\alpha$ emission and UV continuum \citep{Lee2009, Neff2015}; H$\alpha$ emission comes from the recombination of HII gas ionized by the massive O stars, while the UV continuum flux comes from the photospheres of O and B stars themselves. Deep H$\alpha$ imaging has produced a growing census of outlying HII regions in the far outskirts of galaxies, well beyond the galactic disks \citep{Gerhard2002, Ryan-Weber2004, Boquien2007, Werk2010}. Galactic HII regions have a typical SFR(H$\alpha$)/SFR(NUV) ratio in the range of [1-3], corresponding to a range in F(H$\alpha$)/F(NUV) ratio spanning $[1.5-2.0] \ \rm \AA^{-1}$ \citep{Werk2008}. Most outlying/intergalactic HII regions lie within this range or below, exhibiting a UV excess \citep{Boquien2007}. A wide range in F(H$\alpha$)/F(NUV) ratios can be explained by various star formation histories and stellar IMFs \citep{Boquien2007, Werk2010}. The star symbol in Figure \ref{fig:ratios} assumes a SFR(H$\alpha$)/SFR(NUV) ratio equal to 1. The H$\alpha$/UV ratio in the Ribbon gas is too high to be dominated by star formation, though star formation may be contributing to the emission in part. 

\begin{equation}
\label{eq:q0}
Q_{0} = \frac{L(H\alpha)}{h\nu_{H\alpha}}\frac{\alpha_{\beta}}{\alpha_{H\alpha}^{eff}} 
\end{equation}

We estimate the number of ionizing photons $\rm Q_{0}$ required to produce the Ribbon emission following the procedure for an optically thick nebulae given in \cite{Osterbrock2006}. In Equation \ref{eq:q0}, the case B recombination coefficient is $\alpha_{\beta} = 2.59 \times 10^{-13} \rm cm^{3} \ s^{-1}$ (assuming T $=10^{4}$ K, $\rm n_{e} = 1 \ cm^{-3}$) and the H$\alpha$ effective recombination coefficient is $\alpha_{H\alpha}^{eff} = 1.17 \times 10^{-13} \rm cm^{3} \ s^{-1}$ (assuming T $=10^{4}$ K). The average H$\alpha$ luminosity of the Ribbon gas L(H$\alpha) = 8.63\times 10^{38} \rm \ ergs \ s^{-1}$ requires approximately $\rm Q_{0} = 6.33 \times 10^{50} \rm \ ph \ s^{-1}$, the equivalent of $\sim 19$ O5 stars, $\sim 48$ O7 stars, or $\sim 174$ O9 stars (Table 2.3 in \cite{Osterbrock1974}). Using a mass-to-light ratio of 0.02 for a sub-solar metallicity SSP \citep{Bruzel2003}, the Ribbon gas has a stellar mass of $843\rm \ M_{\odot}$ or approximately 17 O stars (assuming an O star mass of $50 \rm \ M_{\odot}$), corresponding to a limiting Vega magnitude of $m = 24$. 

Extended star formation is likely tied to the merger history of the CGM as it may be occurring in tidal debris from past interactions \citep{Werk2008} or in faint dwarf galaxy satellites \citep{Lee2009}. Dwarf galaxies with a SFR of $\rm < 0.003 \ M_{\odot} \ yr^{-1}$, on par with the SFR derived in the Ribbon gas, have an average H$\alpha$/UV ratio that is a factor of two lower than expected in spiral galaxies \citep{Lee2009}. This is inconsistent with the Ribbon gas, which shows enhanced H$\alpha$ emission relative to UV. 

The lower H$\alpha$/UV ratio in the star forming ring and Filament A is closer to the expected ratio for Galactic HII regions, suggesting the gas is primarily excited by young massive stars; however, the H$\alpha$ flux is still slightly higher than expected if produced by young stars alone. The stellar population in the extended star forming ring is resolved with archival HST ACS imaging. 

\begin{deluxetable*}{ccccc}
\tablecaption{CH$\alpha$S Derived Properties}
\label{table:derived}
\tablehead{\colhead{Feature} & \colhead{SFR$(H\alpha)$} & \colhead{$\Sigma_{SFR}$} & \colhead{$\Sigma_{gas}$} & \colhead{$N_{gas}$} \\ & [$M_{\odot} \ \rm yr^{-1}$] & [$M_{\odot} \ \rm yr^{-1} \ kpc^{-2}$] & [$M_{\odot} \ \rm pc^{-2}$] & [$\rm \times 10^{20} \ cm^{-2}$] }
\startdata
Filament A & $ 1.64 \pm 1.30 \times 10^{-2} $ & $ 1.87 \pm 1.49 \times 10^{-3} $ & $ 4.46 \pm 2.32 $ & $ 5.56 \pm 2.89$ \\ 
Ribbon & $ 4.75 \pm 1.57 \times 10^{-3} $ & $ 1.52 \pm 0.50 \times 10^{-3} $ & $ 3.67 \pm 0.85 $ & $ 4.58 \pm 1.07 $ \\ 
SF Ring & $ 5.68 \pm 3.81 \times 10^{-3} $ & $ 2.16 \pm 1.45 \times 10^{-3} $ & $ 4.86 \pm 2.18 $ & $ 6.06 \pm 2.72 $ 
\enddata
\end{deluxetable*}

\subsubsection{Collisional Ionization and Shocks}
Classical evidence for shock structure on large scales, such as thin filamentary structure and line splitting along the minor axis, is tenuous in NGC 1068. The $40^{\circ}$ inclination of the galactic disk makes it difficult to decouple emission from the galactic halo and emission from the disk material overlapping in projection \citep{Brinks1997, Bland-Hawthorn1997, Veilleux2002}. High radial velocity outflows up to $3200$ km s$^{-1}$ traced by line-emitting material are found near the nucleus of NGC 1068 \citep{Cecil2002}; however, UV spectra show these high velocity narrow line regions are not consistent with shocks, and are instead photoionized and radiatively accelerated \citep{Cecil2003}. We consider a number of shock models for the extended Ribbon gas and show that the UV-optical ratios are possibly consistent with a slow shock on the order of 100 km s$^{-1}$ (See \cite{Shull1979} model in Figure \ref{fig:ratios}). 

There are a variety of possible mechanisms for driving shocks in the extended ionized gas surrounding NGC 1068. Extended emission line regions aligned with the axis of a radio jet may be the result of jet-cloud interactions, producing shocks that induce star formation. There are a few examples of jet induced star formation at low redshifts (e.g., \cite{Croft2006, Cresci2015, Capetti2022}). In NGC 5128, a jet-cloud interaction is supported by the detection of a filament of ionized gas and embedded star forming regions stripped from the nearby HI cloud \citep{Schiminovich1994, Joseph2022}; both the ionized gas and the tip of the HI cloud exhibit a similar velocity deviation from the expected rotation velocity around Centaurus A \citep{Oosterloo2005}. In Minkowski's Object a jet-cloud interaction is supported by the knotty/filamentary morphology of the ionized gas, the kinematically disturbed gas velocities, young resolved stellar clusters in HST imaging, and the deflection of the radio jet \citep{VanBreugel1985, VanBreugel1993, Croft2006}. 
The jet-cloud interaction model is particularly appealing in NGC 1068 because it results in a Ribbon feature composed of young stars embedded in warm line-emitting gas. In NGC 1068, the velocity structure of the ionized Ribbon gas as well as bifurcations in the gas morphology may support a jet-cloud interaction. The velocity of the Ribbon gas is consistent with the known HI rotation curve but with a small velocity gradient across the Ribbon on the order of 40 km s$^{-1}$. This velocity gradient is comparable to entrained gas in the jet-cloud interaction around Minkowski's Object \citep{Croft2006} but not as high as the anomalous velocity gradient found around Centaurus A \citep{Oosterloo2005}. Jet-cloud interactions have already been suggested in NGC 1068 close to the central AGN \citep{Gallimore2004}, but have yet to be extrapolated to the extended gas structure. Unlike the inner filametary features around NGC 1068 (Filament A and B), the Ribbon gas is misaligned with the current axis of the central jet by approximately $30^{\circ} - 50^{\circ}$; however, the morphology of the jet on large scales is unknown. The northern lobe of the radio jet is already deflected by dense molecular clouds close to the nucleus and bends abruptly from a position angle of $\sim 12^{\circ}$ to $\sim 30^{\circ}$ \citep{Gallimore1996, Gallimore2004, Garcia-Burillo2014, Mutie2024}. Further deflection on large scales could bring the Ribbon gas and the jet axis into better alignment. There is also evidence to suggest the nucleus of NGC 1068 actually contains two AGN, the result of a minor merger, with the two jet axes slightly offset in position angle \citep{Shin2021}. If jet-induced star formation is taking place at large radii in the ionized Ribbon around NGC 1068, it would be one of very few laboratories for studying this astrophysical phenomenon in the local universe \citep{Oosterloo2005, VanBreugel1993}.

An alternative model, which does not rely on the alignment of the extended emission with respect to the jet-axis, is a galactic wind. A broad galactic wind driven by feedback from massive stars may heat the Ribbon gas via ram-pressure induced shocks or instability-driven turbulence. Starburst-driven winds are poorly collimated; they are capable of ionizing gas over a wide cone angle and can reach all the way to the intergalactic medium \citep{Veilleux2002, Strickland2000}. Many galaxies have extended emission line features at large distances that are thought to be excited by a starburst driven wind (e.g., \cite{Devine1999, Veilleux2003}). This phenomenon is less common in galaxies with AGN but has been observed in galaxies such as NGC 3079 and NGC 5128. An analogous mechanism could be at play in NGC 1068, which hosts an inner ring of galactic starburst activity \citep{Garcia-Burillo2014} and an extended star forming ring beyond the disk \citep{Thilker2007}. 

\subsection{Comparison to other galaxies with EELR/ENLRs}
The detection of extended emission in NGC 1068 invites comparison with a host of other galaxies at low and high redshift that feature such extended emission far beyond the nuclear regions and central disk.  The general picture is one of a luminous, obscured AGN ioinzing clouds present above the ISM, whether in the CGM or in disturbed in-situ or accreted structures resulting from mergers or other tidal encounters. 

Galaxies with similarly bright H$\alpha$ or OIII blobs have been identified through IFU surveys
or dedicated narrowband or broadband searches \citep{Husemann2014, Keel2017, Bait2019, Balmaverde2022, Keel2024}. Hanny's Voorwerp \citep{Lintott2009, Keel2012} is an example of a light echo from a now-fading QSO,
tracing AGN activity on 10s of kyr timescales. Other spectacular nearby examples of AGN with extended ionized gas such as NGC 5972 may also provide evidence of fading or evolution in the AGN ionizing luminosity \citep{Keel2012, Keel2017, Treister2023, Thomas2023}. HII regions in the disk of NGC 1068 are primarily ionized by the AGN \citep{D'Agostino2018} and this disk emission has recently been proposed as a light echo attributed to a burst in AGN activity $\sim 2000$ years ago \citep{Hviding2023}. The Ribbon gas may similarly be an ionized remnant of this burst in AGN activity. Our observation of ultra-extended emission from the Ribbon in NGC 1068 probe levels near the typical limits used for EELR/ENLR size determinations \citep{Balmaverde2022}. The Ribbon gas, star forming ring, and Filament A all likely lie at different absolute radii; therefore, processes operating on the order of the light travel delay between features – such as AGN variability on short timescales – may contribute to differences in their observed properties and complicate any direct comparison amongst them.

One relatively local exemplar of a complex extended emission line region ionized by multiple processes is Coma A/3C277.3 at $\rm z = 0.085$ a bright double-lobed radio source with a supermassive black hole roughly 10 times more massive than NGC 1068 [$10^8$ vs. $10^7$ M$_\odot$] \citep{Capetti2022}.  The extended (60 kpc) emission-line region around Coma A shows clear evidence for widespread and knotty recent star formation, an ionization cone, and evidence for fast shocked gas. Extended emission around the ``Teacup galaxy" at $\rm z\sim 0.1$ is similarly complex including gas photoionized by the AGN, an ionized galactic outflow, and AGN-triggered star formation \citep{Venturi2023}. 

Kinematically disturbed regions with line widths suggesting an outflow are typically found at a smaller radii than the extended narrow line regions (e.g., \cite{Sun2017, Storchi-Bergmann2018, Deconto-Machado2022}). In these systems, the radii of the narrow line regions, the radii of the kinematically disturbed regions, and the outflow velocities are all correlated with AGN luminosity. NGC 1068 has moderate luminosity and in many respects is representative of nearby AGN in this redshift regime (e.g., BASS survey: \cite{Koss2017}). As described above, at the extended distance of the Ribbon, there is no strong evidence for significant kinematic disturbance in the data, particularly at the level seen in Centaurus A.  Further multi-wavelength kinematic mapping of NGC 1068, and even deeper probes of the ionized gas, should allow us to understand how the extended signatures described here are associated with the flows of gas in and around the system.

\section{Summary}
\label{sec:summary}
\begin{enumerate}
    \item We report the discovery of a new extended emission line region around NGC 1068 located well beyond previously known extended H$\alpha$ filamentary structure, at a projected radius of 20.4 kpc from the galactic center. This Ribbon gas has a range in H$\alpha$ surface brightness from $[4-16]$ R, with fainter regions on the order of the sky background.  The surface brightness falls off gradually in the outer wisps of the Ribbon. We leave to future work probing the fainter and more extended gas features (by binning) that may reach lower surface brightness beyond our signal-to-noise cut.  
    
     \item The Ribbon gas is not associated with any galactic structure or known tidal features in the halo of NGC 1068. Unlike Filament A and B closer to the galaxy, the Ribbon gas is slightly misaligned with the current axis of the central jet; however, the morphology of the jet on large scales is unknown. 

     \item The average velocity of the Ribbon gas is consistent with the known HI rotation curve but with a small velocity gradient across the Ribbon on the order of 40 km s$^{-1}$. The velocity gradient is comparable to entrained gas in the jet-cloud interaction around Minkowski's Object but not as high as the anomalous velocity gradient found around Centaurus A. The velocity structure of the Ribbon gas as well as bifurcations in the gas morphology may signify jet-cloud interactions. Turbulent velocities in the Ribbon feature are low but the Ribbon gas also displays a small dispersion gradient.
     
    \item The morphology of extended emission in H$\alpha$ is correlated with extended UV emission around NGC 1068, and the ratio of H$\alpha$ to UV flux is increasing with radius from the galactic center.  H$\alpha$ to UV flux ratios around NGC 1068 are comparable to extended emission line ratios in the halos of NGC 5128, NGC 253, and M82.  
    
    \item The H$\alpha$/UV ratio in the Ribbon gas is not fully explained by either in-situ star formation or nebular emission alone.  It is possibly consistent with a slow shock on the order of 100 km s $^{-1}$; however, the Ribbon gas is most likely composed of young stars embedded in warm line-emitting gas ionized by a mix of processes. Depending on the assumed geometry of the extrapolated ionization cone emanating from the AGN, the Ribbon gas may fall within the ionization bicone. However, the H$\alpha$/UV ratios and UV ratios are not simply explained by a hard photoionizing radiation field. 
    
    \item The lower H$\alpha$/UV ratio in the star forming ring and Filament A suggests the gas is primarily excited by young massive stars; however, the H$\alpha$ flux is too high to be produced by young stars alone. Similar to the Ribbon gas, these features are also likely a composed of young stars embedded in warm line-emitting gas, ionized by a mix of sources including the central AGN, as previously suggested for Filament A and B.  
    
    \item Follow-up observations of the neutral atomic and molecular gas in the filament and Ribbon structures will be important for understanding the origin of the Ribbon gas located in the circumgalactic medium of NGC 1068, the amount of gas available for star-formation, and the level of AGN-induced star formation occurring in these regions. 
\end{enumerate}

\begin{acknowledgments}
The Circumgalactic H$\alpha$ Spectrograph is funded by NSF AST-1407652 (previously ATI). For part of this project duration Nicole Melso was supported with an NSF GRFP under grant number DGE-1644869, and she is currently supported by the Alan Brass Prize Fellowship in Instrumentation and Technology Development. Meghna Sitaram is supported by NASA FINESST 80NSSC22K1603. For part of this project duration B\'{a}rbara Cruvinel Santiago was supported by NASA FINESST 80NSSC19K1419. This data was collected at the MDM Observatory, operated by Dartmouth College, Columbia University, Ohio State University, Ohio University, and the University of Michigan. Many thanks to the MDM Observatory staff Eric Galayda and Tony Negrete for their help setting up these observations and installing CH$\alpha$S on the Hiltner 2.4-meter. Thank you to John Thorstensen and Jules Halpern for their support in scheduling these observations. Thank you to Justin Rupert and Ben Cassese for collecting follow-up long slit spectra with the Ohio State Multi-Object Spectrograph (OSMOS). 
\end{acknowledgments}

\bibliography{squid}{}
\bibliographystyle{aasjournal}

\end{document}